\documentclass[%
aps,
pre,
reprint,
superscriptaddress,
longbibliography,
a4paper,
]{revtex4-1}
\usepackage[english]{babel}
\usepackage{amssymb,amsmath,stmaryrd,array}

\usepackage{graphicx}
\usepackage{subfig}

\graphicspath{{figs/}}

\makeatletter

\newcommand{\pdr}[2]{\frac{\partial #1}{\partial #2}}

\newcommand{\mum}{$\mu$m}

\newcommand{\degc}{$^\circ$C}

 \newcommand{\vc}[1]{\mathbf{#1}}

 \newcommand{\e}{\mathrm{e}}

\makeatother

\begin{document}
\DeclareGraphicsExtensions{.eps,.png,.pdf}
\title{%
Electrically assisted light-induced azimuthal gliding of 
the nematic liquid crystal easy axis on photoaligned substrates
}

\author{A.V.~Dubtsov}
\affiliation{%
Moscow State University of Instrument Engineering and Computer Science,
 Stromynka 20, 107846 Moscow, Russia}

\author{S.V.~Pasechnik}
\affiliation{%
Moscow State University of Instrument Engineering and Computer Science,
 Stromynka 20, 107846 Moscow, Russia}
\affiliation{%
Hong Kong University of Science and Technology, 
Clear Water Bay, Kowloon, Hong Kong}

\author{Alexei~D.~Kiselev}
\email[Email address: ]{kiselev@iop.kiev.ua}
\affiliation{%
 Institute of Physics of National Academy of Sciences of Ukraine, 
prospekt Nauki 46, 03028 Kiev, Ukraine}
\affiliation{%
Hong Kong University of Science and Technology, 
Clear Water Bay, Kowloon, Hong Kong}

\author{D.V.~Shmeliova}
\affiliation{%
Moscow State University of Instrument Engineering and Computer Science,
 Stromynka 20, 107846 Moscow, Russia}

\author{V.G.~Chigrinov}
\email[Email address: ]{eechigr@ust.hk}
\affiliation{%
Hong Kong University of Science and Technology,
Clear Water Bay, Kowloon,
Hong Kong}

\date{\today}

\begin{abstract}
 We study azimuthal gliding of the easy axis
that occurs in nematic liquid crystals
brought in contact with the photoaligned substrate
(initially irradiated azo-dye film)
 under the action of reorienting UV light combined with 
in-plane electric field.
For irradiation with the linearly polarized light, 
dynamics of easy axis reorientation
is found to be faster as compared to the case of non-polarized light.
Another effect is that it slows down with the initial irradiation dose used to prepare the
azo-dye film.
This effect is interpreted  by using the previously suggested phenomenological model.
We present the theoretical results  
computed by solving the torque balance equations of 
the model that agree very well with the experimental data.
\end{abstract}

 \maketitle

\section{Introduction}
\label{sec:intro}

It is well known that
irradiation of a photosensitive (dye or polymer) layer
with linearly polarized UV (LPUV) or visible light
may have a profound effect on its properties 
by producing anisotropy of the angular distribution of molecules. 
In a liquid crystal (LC) brought in contact with the irradiated layer, 
the surface ordering originated from the photoinduced anisotropy
determines the anchoring characteristics such as 
the anchoring strengths and the \textit{easy axis},
$\vc{n}_e$, directed along the direction of
preferential orientation of LC molecules at the surface. 

So, in this way, the light can be used as a 
means to control the anchoring characteristics of photosensitive materials.
Technologically, this idea
underlies the photoalignment (PA) technique 
which is employed in the manufacturing process of liquid crystal displays 
for fabricating high quality aligining substrates~\cite{Chigrin:bk:2008}. 
In the PA method, the easy axis is determined by 
the polarization of the pumping UV light,
whereas the azimuthal and polar anchoring
strengths may depend on a number of the governing parameters such as
the wavelength and the irradiation dose~\cite{Kis:pre2:2005}.  

Macroscopically, overage orientation of LC molecules
in interfacial layers is 
described by the \textit{surface director}, $\vc{n}_s$,
which is paralell to the easy axis at the equilibrium spatially uniform
state.  The latter is no longer the case
when the orientational structure is
deformed by external (electric and/or magnetic) fields.
So, in the presence of field-induced director deformations,
the surface director may deviate from the easy axis. 
 
It turned out that the easy axis itself may 
slowly rotate under the action of strong (magnetic or electric) fields.
Over the past few decades this slow motion~---~the so-called 
\textit{easy axis gliding}~--~has received much 
attention as a widespread phenomenon
observed in a variety of liquid crystals
on amorphous glass~\cite{Oliveira:pra:1991},
polymer~\cite{Vetter:jjap:1993,Vorflusev:apl:1997,Faetti:epjb:1999,Janossy:pre:2004,
Joly:pre:2004,Faetti:pre:2005,Faetti:lc:2006,Pasechnik:lc:2006,Buluy:jsid:2006,Janossy:pre:2010}   
and solid~\cite{Faetti:epjb:1999,Pasechnik:lc:2006} 
substrates. 
 
Recently, it was found that, in addition to magnetically (electrically) induced gliding,
there is the photoinduced gliding of the easy axis in dye-doped liquid
crystals~\cite{Statman:lc:2008,Fedorenko:pre:2008}.
According to Ref.~\cite{Fedorenko:pre:2008},
this effect is driven by the processes of light-induced adsorption/desorption 
of dye molecules on an aligning polymer surface.

Slow reorientation of the easy axis
also takes place on the photosensitive layers prepared using 
the PA technique~\cite{Buluy:jsid:2006,Pasechnik:lc:2006}.
So, in a LC cell with the initially irradiated layer, 
subsequent illumination with reorienting light which polarization
differs from the one used to prepare the layer
can trigger the light-induced easy axis gliding. 
Such gliding  may be of considerable interest 
for applications such as LC rewritable devices~\cite{Chig:jjap:2008}.

In our recent paper~\cite{Pasechnik:lc:2008} 
we have found that the photoinduced azimuthal gliding of the easy axis
on a photoaligned substrate
may occur when 
irradiation with linearly polarized UV light (LPUV)
is used in combination with an ``in-plane'' ac electric field
($f=3$~kHz) 
produced by the voltage applied in the lateral direction. 
This effect might be called the
\textit{electrically assisted light-induced azimuthal gliding
of the easy easy}. 

In Ref.~\cite{Pasechnik:lc:2008}, 
the emphasis was on slow relaxation of the easy axis 
back to the initial state after turning off both 
the electric field and light. 
This stage of switching off relaxation was
controlled by the four parameters: 
the electric field strength $E$, the light intensity $I$, 
the exposure time $t_{exp}$ and the doze $D_p$ of the initial UV irradiation.
It was found that, at certain
combinations of the governing parameters, 
relaxation considerably slows down (up to several days). 
 
So, the combined effect may 
be used  to tune technical parameters of LC memory devices.
Similarly, in the earlier paper~\cite{Cui:lc:1999},
an assisting in-plane electric field has been used
as a stimulating factor of the light induced homeotropic-to-planar 
transition to improve the response sensitivity of the photo-driven cells.
In addition, 
it has to be taken into account in applications of LC and
PA technique using different field combinations that involve 
intense light beams along with strong electric fields
(e.g. in photonics~\cite{Pasechnik:bk:2009}).

In this study azimuthal gliding of the easy axis in the regime of switching on
both the in-plane electric field and reorienting 
(linearly polarized and non-polarized)
UV  light (LPUV and NPUV)
will be of our primary interest.
The layout of the paper is as follows.

In Sec.~\ref{sec:experiment}, we describe the
experimental procedure and setup used to measure 
the azimuthal easy axis angle as
a function of time.
The experimental results 
for the azo-dye films prepared at different initial
irradiation doses, $D_p$, 
are presented in Sec.~\ref{sec:results}.
In particular, we find that,
in the presence of the assisting in-plane electric field,
LPUV and NPUV irradiation
may equally produce slow changes in LC surface orientation.
Another effect is that easy axis reorientation dynamics 
slows down with the initial irradiation dose, $D_p$.
In Sec.~\ref{subsec:model},
we apply the phenomenological model of azimuthal
gliding~\cite{Pasechnik:lc:2008}
to interpret the experimental data measured in the cells
with the azo dye layer prepared at different values of $D_p$.
Finally, in Sec.~\ref{sec:conclusion} we discuss the results
and make some concluding remarks.

\begin{figure}
\centering
\resizebox{90mm}{!}{\includegraphics*{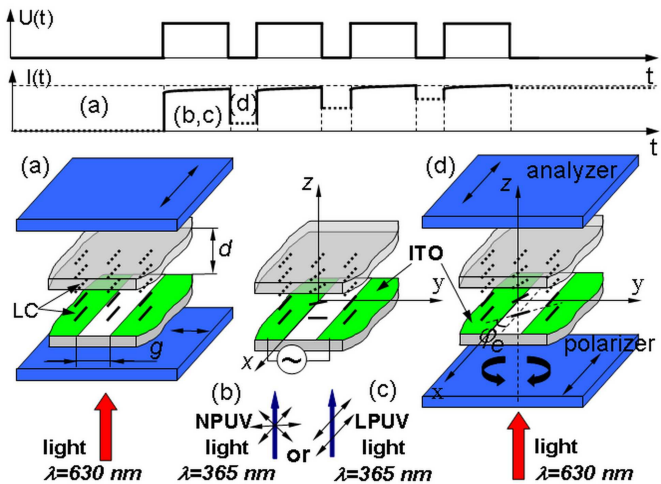}}
\caption{%
Geometry of the experiment: 
(a) initial state; 
(b) switching on an in-plane electric field  combined with NPUV irradiation; 
(c)   switching on an in-plane electric field  combined with LPUV irradiation; 
(d) switching off the electric field and light to 
measure the easy axis azimuthal angle $\phi_e$.
}
\label{fig:expt_geom}
\end{figure}

\section{Experiment}
\label{sec:experiment}

For our measurements, two differently treated glass plates were
assembled to prepare the planar LC cell. 
The cell thickness was $d=17.4 \pm 0.2$~\mum. 
The upper glass plate with a rubbed polyimide film  
was to provide the strong planar anchoring conditions. 
The lower one with transparent ITO electrodes and
the interelectrode stripes (the gap was $g=50$~\mum) was covered 
with a film of photoaligning substance, the azobenzene sulfric dye SD1 (Dainippon Ink and
Chemicals)~\cite{Chigrin:bk:2008}. 
The surface of the azo-dye film  was irradiated with
linearly polarized UV light ($\lambda = 365$~nm) 
so as to produce zones illuminated at
different energy exposure doses $D_p$ ranged from $0.07$~J/cm$^2$
to $0.55$~J/cm$^2$. 
In this way, there were typically four zones 
that differ from one another in the initial exposure dose, $D_p$,
and thus in the initial azimuthal anchoring strength,  $W_0$. 
The nematic liquid crystal mixture E7 was injected into
the cell in an isotropic phase by capillary action and 
then slowly cooled  to room temperature.
Note that similar experimental procedure has already been used 
in our previous study~\cite{Pasechnik:lc:2008} .   

Experimental set-up was based on a polarized microscope connected with
a fiber optics spectrometer and a digital camera, where rotating
polarizer technique was used to measure the azimuthal angle
$\phi_e$ characterizing orientation of the easy axis. 
The measurements were carried out at a temperature of 26~\degc.

We used the irradiation of a UV lamp of 6W power at the wavelength 
$\lambda= 365$~nm 
(absorption spectrum for SD1 has a peak at 365~nm)
as a reorienting light beam which, 
as is shown in Fig.~\ref{fig:expt_geom},
was normally impinging onto the bottom substrate coated with the azo-dye layer. 
Referring to Fig.~\ref{fig:expt_geom},
for LPUV irradiation, the polarization plane was directed along
the initial orientation  of LC molecules, $\vc{n}_0$, 
which is parallel to the electrode stripes. 
So, LPUV light and the  electric field
make the LC molecules align in the same direction. 

Each measurement was conducted on a fresh area of the cell, 
protected from the negative irradiation.  
In order to register microscopic images and 
to measure the value of the easy axis  azimuthal angle, $\phi_e$,
the electric field and light were switched off for about 1 min.
This time interval is short enough 
(dynamics at surface is  considerably slower than that in the bulk)
to ensure that orientation of the easy axis remains essentially intact
in the course of measurements.

\begin{figure}
\centering
   \resizebox{90mm}{!}{\includegraphics*{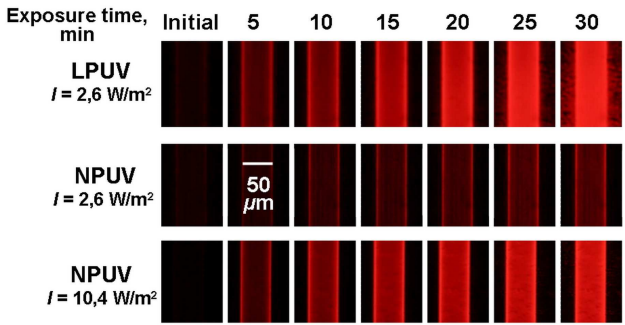}}
\caption{%
Microscopic images of the cell (filter with $\lambda =630$~nm was used)
in crossed polarizers for LPUV and NPUV irradiation at different
exposure times. 
The interelectrode gap ($g=50$~\mum) is indicated.
The ac electric field is $E=2$~V/\mum\,
and the initial irradiation dose is $D_p=0.27$~J/cm$^2$.
}
\label{fig:expt_img}
\end{figure}

\section{Results}
\label{sec:results}

Figure~\ref{fig:expt_img} shows the microscopic images
obtained at different exposure times for irradiaion 
with LPUV and NPUV reorieting light.
In this case, the in-plane ac voltage ($U=100$~V, $f=3$~kHz) and the initial irradiation dose
($D_p=0.27$~J/cm$^2$) are kept constant.  
Note that, when applying sufficiently strong electric field ($E=2$~V/\mum) 
in combination with reorienting irradiation for more than 30 minutes,
memory effects take place. 
It means that, after switching off the field and light,
the easy axis did not relax back to its initial state for at least few months.

When an in-plane ac electric field 
is applied to the transparent ITO electrodes, 
twist-like deformations of LC director occur. 
In crossed polarizers,  
such deformations reveal themselfves as
the bright stripes that can be seen in Fig.~\ref{fig:expt_img}. 

\begin{figure}
\centering
   \resizebox{90mm}{!}{\includegraphics*{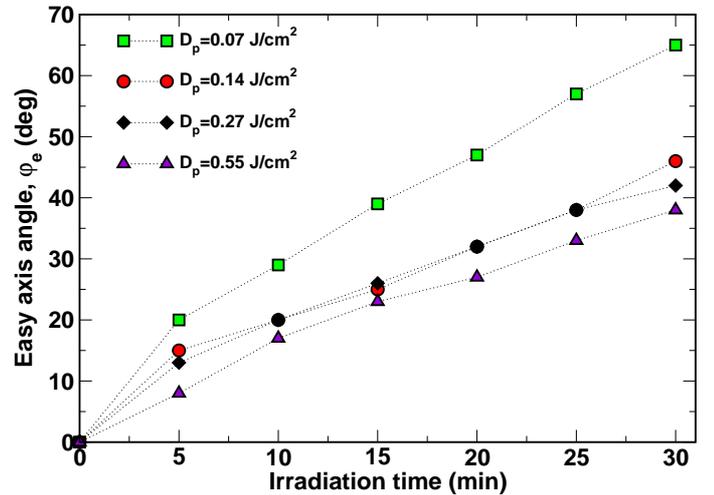}}
\caption{%
Easy axis angle as a function of the irradiation time
measured in the samples prepared at different initial irradiation doses $D_p$.
The reorienting  UV light
is nonpolarized with $\lambda=365$~nm and $I=10.4$~J/m$^2$;
the applied voltage is $U=100$~V.
}
\label{fig:nonpol}
\end{figure}

Given the light intensity $I=2.6$~W/m$^2$,
reoriention dynamics at LPUV irradiation is 
considerably faster as compared to 
the case of NPUV light. 
In order to produce variations of optical picture 
similar to those for LPUV light,
the intensity of NPUV irradition had to
be increased at least in four times from 2.6~W/m$^2$ to 
10.4~W/m$^2$.
This can be seen from Fig.~\ref{fig:nonpol}
and Fig.~\ref{fig:lin}, where
the easy axis angle $\phi_e$
is plotted against the irradiation time
at different initial irradiation doses $D_p$
for NPUV and LPUV light, respectively.

According to Fig.~\ref{fig:nonpol}
and Fig.~\ref{fig:lin}, reorientation of the easy axis 
slows down with the initial irradiation dose $D_p$
that determines the initial value of the azimuthal anchoring strength. 
In~\cite{Kis:pre2:2005}, this anchoring strength was 
found to be an increasing function of the dose $D_p$.
So,  the lower the initial azimuthal anchoring strength
the faster gliding is.

\subsection{Model}
\label{subsec:model}

We can now apply the phenomenological model
formulated in~\cite{Pasechnik:lc:2006,Pasechnik:lc:2008}
to describe this effect.
According to this model,
the anchoring characteristics of the photoaligned layer
such as the easy axis, $\vc{n}_e$,
are determined by orientational order of 
LC molecules adsorbed by the azo-dye film. 
The adsorbed molecules can be divided into two ensembles
characterized by two mutually orthogonal 
directions of preferential orientation of long molecular axes: 
$\vc{n}_0$ and $\vc{n}_2$, $\vc{n}_0\perp\vc{n}_2$. 
Interestingly, this assumption bears a resemblance to the dual-axis models
previously employed 
to study competetive effects of photoalignment and microgrooves~\cite{Osipov:jap:2002}
and to describe
the anchoring transitions on substrates irradiated with light and a plasma 
beam~\cite{Andrien:jetp:1997,Kis:pre:2008}.

Ordering of the first ensemble 
is governed by the orientational order parameter of the azo dye layer
induced by the initial irradiation with linearly polarized UV light
(PA treatment made before the cell is filled with LC).
The direction of preferential orientation, $\vc{n}_0$, is dictated by
the azo-dye order parameter which is known to depend on 
the irradiation dose $D_p$~\cite{Kis:pre2:2005}.
So, the initial number of LC molecules 
in this ensemble $N_0(0)$ and the initial anchoring strength $W_0(0)$ 
are determined by the initial irradiation dose $D_p$.  

When the cell filled with LC is irradiated
by the reorienting light with
polarization normal to the one used
at the preparation stage for PA treatment of
the azo-dye layer,
absorbed LC molecules 
undergo light-induced transitions to the second ensemble
aligned along the vector, $\vc{n}_2$, orthogonal to $\vc{n}_0$.
These transitions occur due to
 reorientation of dye molecules interacting with the adsorbed LC layer.
 Since the anchoring strengths, $W_0$ and
 $W_2$, 
for the easy axes of the ensembles, $\vc{n}_0$  and $\vc{n}_2$,
 are proportional to the corresponding numbers of molecules $N_0$ and
 $N_2$:
$W_{0,\,2}\propto N_{0,\,2}$,
it can be concluded that the strength $W_0$ ($W_2$)
decreases (increases) with irradiation time.

 The second ensemble can also be populated by applying 
 an in-plane electric field in the direction $\vc{n}_2$.   
It can be explained in terms
of adsorption-desorption processes taking place in the near-surface
layer of characteristic thickness $h$. 
Initially, in the absence of electric field,
the surface director $\vc{n}_s$ characterizing
average orientation of LC molecules  in the near-surface layer 
is directed along the easy axis  $\vc{n}_0$. 
So, the adsorption-desorption
processes do not influence
the undisturbed angular distributions of LC molecules
in the absorbed and near-surface layers
which are initially identical. 

 An electric field $\vc{E}$ produces a twist deformation on the
 distance $\xi$ defined as the electric coherence length: 
 \begin{equation}
   \label{eq:coh_length}
   \xi=\frac{1}{E}\sqrt{K_{22}/(\varepsilon_0\Delta\varepsilon)},
 \end{equation}
where $\Delta\varepsilon$ is 
the electric permittivity anisotropy, $K_{22}$ is the Frank elastic
 constant for the twist deformation. 
The distance $\xi$ decreases with $E$ down to
 the values comparable with the thickness of the near-surface layer $h$. 
In our case,
for the liquid crystal mixture E7 with the twist elastic constant
$K_{22}\approx 6.5\times 10^{-12}$~N
and the dielectric anisotropy $\Delta\varepsilon\approx 13.7$,
the electric coherence length $\xi$ 
can be estimated at about $0.12$~\mum.

\begin{figure}
\centering
\resizebox{90mm}{!}{\includegraphics*{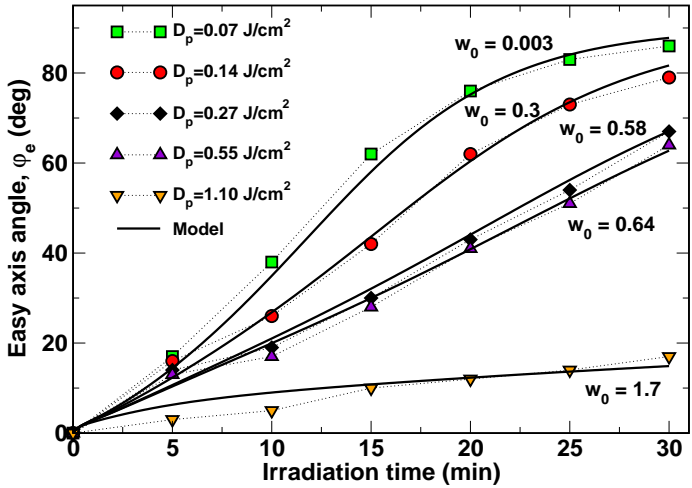}}
\caption{%
Easy axis angle versus the irradiation time
at different initial irradiation doses $D_p$.
The reorienting  UV light
is linearly polarized 
with $\lambda=365$~nm and $I=2.6$~J/m$^2$;
the applied voltage is
$U=100$~V.
Solid lines represent the theoretical curves
computed 
at the indicated initial values of 
the anchoring parameter $w_0=W_0(0)(\xi-h)/(2K_{22})$.
}
\label{fig:lin}
\end{figure}

Under this electic-field-induced deformation, the surface director,
$\vc{n}_s$, inclines towards the direction $\vc{n}_2$ 
($\vc{E}\parallel \vc{n}_2$ and $\Delta\varepsilon>0$).
The absorption-desorption processes involving exchange of molecules 
between the differently aligned (absorbed and near-surface) layers
will eventually produce reorientation of absorbed molecules 
along the direction of the electric field.
Note that, owing to low probability of adsorption-desorption events,  
noticeable changes may require very long periods of time.
In terms of the model, 
this can be described an increase in the number of molecules $N_2$ 
and in the anchoring strength $W_2$.

Though the mechanisms behind slow reorientation of adsorbed LC molecules
 induced by light and electric field 
are essentially different,
in the phenomenological model~\cite{Pasechnik:lc:2006,Pasechnik:lc:2008}, 
they can be treated along similar lines.
In particular, orientation of the easy axis $\vc{n}_e$ characterized by the azimuthal
 angle $\varphi_e$ is defined by the balance of 
the two torques arising from the two
 ensembles $N_0$ and $N_2$ : the torque transmitted from the bulk by the near-surface layer 
and the viscous torque proportional to \textit{the specific viscosity of gliding} $\gamma_e$. 
For the surface director $\vc{n}_s$ with the azimuthal angle 
$\varphi_s$,
the analogues balance involves the torque arising 
due to deviation of the surface director from the easy
axes (it is proportional to the surface anchoring energy strength
$W_s$),  the torque transmitted from the bulk and the viscous torque
proportional to the \textit{surface viscosity} $\gamma_s$. 
The resulting system of equations
 for the easy axis and surface director azimuthal angles, $\varphi_e$ and $\varphi_s$,
was derived in~\cite{Pasechnik:lc:2008}  
and can be conveniently written 
in the following dimensionless form: 
\begin{subequations}
  \label{eq:system}
\begin{align}
&
  \label{eq:phi_e}
\pdr{\varphi_e}{\tau}=
  [\pi/2-\varphi_s] -
w_e \sin 2\varphi_e,
\quad
\tau=t/\tau_e
\\
&
 \label{eq:phi_s}
\pdr{\varphi_s}{\tau}=
\gamma
\Bigl\{  
[\pi/2-\varphi_s] -
w_s \sin 2(\varphi_e-\varphi_s)
\Bigr\},
\end{align}
\end{subequations}
where $\tau_e=\gamma_e(\xi-h)/K_{22}$ 
is the characteristic time  of 
easy axis reorientation;
$\gamma=\gamma_e/\gamma_s$ 
is the viscosity ratio;
$w_{e,\,s}=W_{e,\,s}(\xi-h)/(2K_{22})$ is
the dimensionless anchoring parameter
and 
$W_e=W_0-W_2$ is the effective anchoring parameter which defines 
the strength of coupling between
the easy axis $\vc{n}_e$ and
the initial state of surface orientation 
described by the vector $\vc{n}_0$.

From the above discussion,
the initial value of the easy axis anchoring parameter (coupling
strength)
$W_e$ is determined by the initial irradiation doze $D_p$ and
equals $W_0(0)$.
In the course of irradiation with reorienting light assisted by 
an in-plane electric field,
the anchoring parameter $W_e$ declines and can even become negative.
The latter indicates that  
the direction of preferential orientation at the azo-dye film
is changed from $\vc{n}_0$ to $\vc{n}_2$.  

For the purposes of modelling, the exponential dependence
 \begin{align}
   \label{eq:W-e}
   W_e=W_0(0)-A_L (1-\e^{-t/\tau_L})-A_E (1-\e^{-t/\tau_E}),
 \end{align}
where the characteristic times $\tau_L$ ($\tau_E$)
and the amplitudes $A_L$ ($A_E$)
 are related to the action of light (electric field),
can be regarded as a reasonable approximation.

\begin{figure}
\centering
   \resizebox{90mm}{!}{\includegraphics*{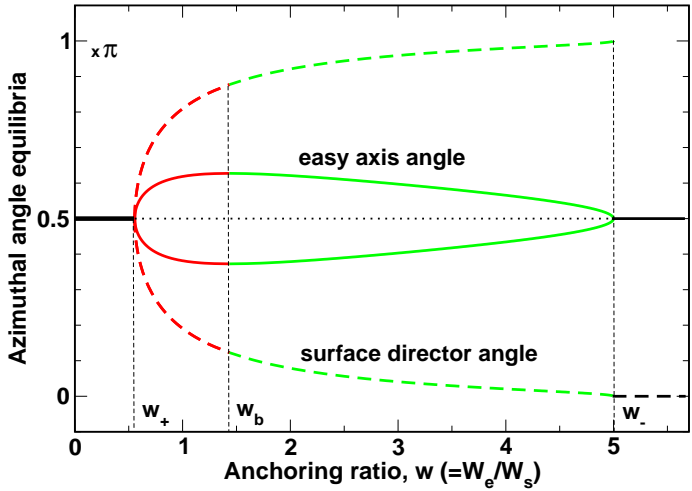}}
\caption{%
Bifurcation diagram for equilibria
of easy axis and surface director azimuthal angles
computed from Eq.~\eqref{eq:bifur-curves}
as a function of the anchoring ratio, 
$w=W_e/W_s$, at $w_s=0.4$.
}
\label{fig:bifur}
\end{figure}
 
Before making comparison between the model and experiment, 
we shall dwell briefly on the general properties of 
the dynamical system~\eqref{eq:system}.
Since
the gliding viscosity $\gamma_e$ is typically several orders higher
than the surface viscosity $\gamma_s$,
the viscosity ratio is large, $\gamma\gg 1$. 
In this case,  in the initial stage of reorientation process, 
the surface director angle, $\varphi_s$, 
changes abruptly reaching the adiabatic regime.
In this regime,
the temporary evolution of the angles is subject to the linking relation:
\begin{align}
  \label{eq:adiab-rel}
  \pi/2-\varphi_s=
w_s \sin 2(\varphi_e-\varphi_s).
\end{align}
So, the easy axis angle,   $\varphi_e$,
can be conveniently expressed in terms of the surface
director angle, $\varphi_s$, as follows:
\begin{align}
  \label{eq:adiab-linking}
  \varphi_e =
\pi k_{\nu}+\dfrac{\pi}{4}\, (\nu+1)-
\Bigl(
\psi_s + \dfrac{\nu}{2}\,\arcsin[\psi_s/w_s]
\Bigr),
\end{align}
where $\psi_s=\pi/2-\varphi_s$,
$\nu=\pm 1$ and $k_{\nu}\in \mathbb{Z}$ is the integer.
Then substituting the relation~\eqref{eq:adiab-linking}
into Eq.~\eqref{eq:phi_e} gives
the dynamic equation:
\begin{align}
&
  \label{eq:adiab-dyn-eq}
  \pdr{\psi_s}{\tau}=
\frac{\sqrt{w_s^2-\psi_s^2}}{\sqrt{w_s^2-\psi_s^2}+\nu/2}\,
F_{\nu}(\psi_s),
\\
&
\label{eq:F-nu}
F_{\nu}=
w \left(
\psi_s\cos 2\psi_s +\nu\sqrt{w_s^2-\psi_s^2}\,\sin 2\psi_s
\right)-\psi_s,
\end{align}
where $w=W_e/W_s$ is the \textit{anchoring ratio}. 

The equilibrium states characterized by
the azimuthal angles, $\varphi_e^{(st)}$ and $\varphi_s^{(st)}$,
can be found as 
the stable (attracting) stationary points of the dynamical
system~\eqref{eq:system}.
 It is not difficult to show that, at $|w_s|<0.5$, the surface director
equilibria are among the zeros of the function, $F_{\nu}$, given in
Eq.~\eqref{eq:F-nu}.
More specifically, for $\psi_s^{(st)}=\pi/2-\varphi_s^{(st)}$, we have
\begin{align}
  \label{eq:equil}
  F_{\nu}(\psi_s^{(st)})=0,
\quad
\nu F_{\nu}'(\psi_s^{(st)})<0.
\end{align}
The equilibrium conditions~\eqref{eq:equil} can now be combined with
the expression~\eqref{eq:F-nu} to yield the dependence
of the surface director angle equilibria on the anchoring ratio,
$w=W_e/W_s$,
in the following parametrized form:
\begin{align}
  \label{eq:bifur-curves}
  \varphi_s^{(st)}(w)=
      \begin{cases}
        w\equiv w_{\nu}(t),\\
        \varphi_s^{(st)}=\pi/2 + t, \quad -|w_s|\le t\le |w_s|,
      \end{cases}
\end{align}
where $w_{\nu}(t)=\bigl[\,\cos 2t +\nu\sqrt{w_s^2-t^2}\,\sin(2t)/t\,\bigr]^{-1}$.

Figure~\ref{fig:bifur} shows
the bifurcation curves that are evaluated
using Eq.~\eqref{eq:bifur-curves} and
describe dynamical behaviour of the model 
in relation the anchoring ratio.  

It is seen that there is the only stationary value
of the surface director angle, $\varphi_s^{(st)}=\pi/2$,
provided that $w<w_{+}$ or $w>w_{-}$, where 
$w_{\pm}=w_{\pm}(0)=1/(1\pm 2 w_s)$ is the critical (bifurcation) value of the anchoring ratio.
The equilibrium values of the easy axis angle in these two regions are:
$\varphi_e^{(st)}=\pi/2$ at $w<w_{+}$ and $\varphi_e^{(st)}=0$ at
$w>w_{-}$.
 It means that, 
in the case of sufficiently large coupling parameter $W_e$,
where $w>w_{-}$, the easy axis gliding is completely suppressed,
whereas, in the regime of weak coupling with $w<w_{+}$,
the easy axis rotates  approaching   the stationary state of the
surface director, $\vc{n}_s^{(st)}=\vc{n}_2$.

As it can be seen in Fig.~\ref{fig:bifur},
when the anchoring ratio $w$ passes through the critical points
$w_{\pm}$ the pitchfork bifurcations~\cite{Gucken:bk:1990} occur 
and  the stationary state $\varphi_{s}^{(st)}=\pi/2$
becomes unstable.
So, for the surface director angle in the intermediate region with
 $w_{+}<w<w_{-}$, 
there are two symmetrically arranged stable 
stationary points: $\varphi_{s}^{(\pm)}=\pi/2\pm\Delta\varphi_s^{(st)}$.  
In this case, gliding is not suppressed, but,
by contrast to the regime of weak coupling,
the equilibrium states of the easy axis and the surface director 
are no longer identical.  

Referring to Fig.~\ref{fig:bifur},
there is also the branching point, $w_p=w_{\pm}(w_s)=1/\cos(2w_s)$, 
in this region. At $w<w_p$ ($w>w_p$), the equilibria are defined
by the zeros of the function $F_{+}$ ($F_{-}$).
Note that  the magnitude of the equilibrium angle $\psi_{s}^{st}$
reaches its maximum, $|\psi_{s}^{st}|=|w_s|$, at $w=w_p$. 
In the vicinity of these extrema, relaxation will slow down.
It follows from the expression on the right hand side of dynamical equation~\eqref{eq:adiab-dyn-eq} 
which is proportional to $\sqrt{w_s^2-\psi_s^2}$.

In order to apply the model to interpret experimental data,
the system~\eqref{eq:system} has to be solved numerically.
In Fig.~\ref{fig:lin}, it is demonstrated that by using the model 
we can obtain the theoretical results which are in
good agreement with the experimental data.

The parameters used for computing the curves are:
$\tau_e=10$~min, $\gamma=100$ and $w_s=0.45$.
Since $K_{22}/\xi\approx 5.4\times 10^{-5}$~J/m$^2$,
the azimuthal anchoring strength
$W_s$ can be estimated at $2.4\times 10^{-5}$~J/m$^2$.
For the easy axis anchoring parameter 
$w_e$, we assumed the simplified exponential dependence
$w_e=w_0 - \Delta w (1-\e^{-\tau/\tau_w})$
with $\Delta w  = 1$ and $\tau_w=2$.
The only parameter that depends on the initial dose is 
the initial value of the easy axis anchoring parameter 
$w_0=w_e(0)$.
The values of $w_0$ are shown in Fig.~\ref{fig:lin}:
$w_0=0.003$ ($W_0\approx 1.6\times 10^{-7}$~J/m$^2$) 
at $D_p=0.07$~J/cm$^2$;
$w_0=0.3$ ($W_0\approx 1.6\times 10^{-5}$~J/m$^2$) 
at $D_p=0.14$~J/cm$^2$;
$w_0=0.58$ ($W_0\approx 3.1\times 10^{-5}$~J/m$^2$)  
at $D_p=0.27$~J/cm$^2$;
$w_0=0.64$ ($W_0\approx 3.5\times 10^{-5}$~J/m$^2$) 
at $D_p=0.55$~J/cm$^2$;
and
$w_0=1.7$ ($W_0\approx 9.2\times 10^{-5}$~J/m$^2$)  
at $D_p=1.1$~J/cm$^2$.
So, typical azimuthal anchoring strengths  and
the above estimates for $W_s$ and $W_0$ are of the same order.
As is expected, the initial easy axis anchoring parameter $W_0$ 
increases with the dose $D_p$. 
Interestingly, at $D_p=0.55$~J/cm$^2$, the estimate for the characteristic time
$\gamma_e/W_0=\tau_e/w_0\approx 937$~s is close
to the value, $\gamma_e/W_0\approx 880$~s, reported
in Ref.\cite{Pasechnik:lc:2008}
for the liquid crystal mixture ZhK~616.

\section{Discussion and conclusions}
\label{sec:conclusion}

In this paper we have studied electrically assisted photoinduced
gliding of the NLC easy axis on the photoaligned azo-dye layer.
Our experimental results indicate that, in the presence of 
sufficiently strong in-plane electric field,  
LPUV and NPUV light can both be effective in reorienting the easy axis. 
But, for NPUV irradiation,  the intensity of the pumping beam should be several times 
higher than the intensity of LPUV light.  

We have also found that reorientation dynamics of the easy axis 
depends on the initial dose of UV irradition used to prepare the azo-dye film.
Reorientation is retarded appreciably as the dose increases. 
So, it comes as no surprise that 
the initial irradiation counteracts easy axis gliding
induced by the secondary irradiation with differently polarized
reorienting light.

It is shown that this effect can be described by using 
the phenomenological model previously suggested in 
Refs.~\cite{Pasechnik:lc:2006,Pasechnik:lc:2008}.
The model was formulated as the dynamical system
of the balance torque equations~\eqref{eq:system}
for the easy axis and the surface director azimuthal angles.
Its dynamical behaviour governed by the ratio
of the easy axis and surface director anchoring parameters,
$w=W_e/W_s$, is characterized by two pitchfork bifurcations.
The bifurcation points,$w_{+}$ and $w_{-}$, are at the boundaries of 
the regions with $w<w_{+}$ and $w>w_{-}$
where dynamics of easy axis reorientation occurs in 
the regimes of weak and strong coupling, respectively.
At $w<w_{+}$ the easy axis rotates towards the equilibrium state
with $\varphi_e^{(st)}=\varphi_s^{(st)}=\pi/2$,
whereas, at $w>w_{-}$, reorientation is suppressed and
$\vc{n}_e^{(st)}=\vc{n}_0$
($\varphi_e^{(st)}=0$).

We have fitted the experimental data
by numerically solving the system~\eqref{eq:system}. 
Despite the fact that good agreement between the computed curves
and the data (see Fig.~\ref{fig:lin}) counts in favour of the model,
its theoretical justification has yet to be done. 
In particular, the model in its present form 
cannot be applied 
to the case where the reorienting light is non-polarized.
In order to analyze this case,
at least out-of-plane reorientation has to be taken into 
consideration.

A deeper insight into complicated
molecular mechanisms behind the gliding
under consideration 
requires additional experimental studies and a 
more systematic theoretical treatment 
of the processes involved.
Such treatment
has to deal with the interplay of photoinduced ordering
in azo-dye films~\cite{Kis:pre:2009} 
and the adsorption-desorption 
processes~\cite{Vetter:jjap:1993,Ouskova:pre:2001,Barbero:bk:2006,Fedorenko:pre:2008}
underlying the gliding phenomenon.

This work was partially supported by grants: Development of the Higher
School’s Scientific Potential 2.1.1/5873; Grant NK-410P; HKUST CERG
RPC07/08.EG01 and CERG 612208.


%

\end{document}